\documentclass[a4paper,12pt,fleqn]{article}
\usepackage{amsmath,graphicx}

\newcommand{\pni}{\par\noindent}
\begin{document}
\title{Schwarzschild metrics and quasi-universes}
\author{ A. G. Agnese\footnote{Email: agnese@ge.infn.it} \, and
M. La Camera\footnote{Email: lacamera@ge.infn.it}} 
\date{}
\maketitle 
\begin{center}
{\em Dipartimento di Fisica dell'Universit\`a di 
Genova\\Istituto Nazionale di Fisica Nucleare,Sezione di 
Genova\\Via Dodecaneso 33, 16146 Genova, Italy}\\
\end{center} 
\bigskip
\begin{abstract}\pni
The exterior and interior Schwarzschild solutions are rewritten 
replacing the usual radial variable with an angular one. This 
allows to obtain some results otherwise less apparent or even
hidden in  other coordinate systems.
\end{abstract}
\bigskip \pni
\vspace{1in}\pni
\newpage
\baselineskip = 2\baselineskip
\pni\emph{Introduction}: \quad
It is well known [1] that the three-dimensional space
\begin{equation}
d^{(3)}s^2 = 
\dfrac{dr^2}{1-\dfrac{r^2}{R^2}}+r^2(d\vartheta^2+\sin^2\vartheta
d\varphi^2)
\end{equation}  
of some models of closed homogeneous and isotropic universes has 
an especially simple geometry which can be seen best introducing 
a new angular coordinate $0 \leq \chi \leq \pi$ via $r=R 
\sin\chi$ and transforming the line element (1) into the form 
\begin{equation}
d^{(3)}s^2 = R^2\,(d\chi^2 + \sin^2\chi\, d\Omega^2)
\end{equation}
where
\begin{equation}
d\Omega^2 = d\vartheta^2+\sin^2\vartheta
d\varphi^2 
\end{equation}
The metric (2) is that of a three-dimensional hypersurface of 
radius $R$ which can be represented in a flat, four-dimensional 
Euclidean embedding space.\pni
Our purpose is to employ a similar angular variable to describe 
the geometry of the exterior Schwarzschild solution and to 
investigate such a description of the interior solution also when
$\chi > \pi/2$, a possibility which appears to have been ignored 
in the literature.
\pni\emph{The exterior Schwarzschild solution}: 
The exterior spherically symmetric vacuum solution, which by 
Birkhoff's theorem is also static, will be written in standard 
coordinates as 
\begin{equation}
ds^2 =\dfrac{dr^2}{1-\dfrac{2m}{r}} + r^2 d\Omega^2 -\, N^2(t) 
\left(1-\dfrac{2m}{r}\right) dt^2
\end{equation}
The term $N^2(t)$ allows the matching between exterior and 
interior values of $g_{tt}$ when the interior solution is not 
static and the observer is below the radius $r_1$ of the body; 
of course in the static cases $N^2(t)$ reduces to a constant. 
Such a constant shall be written as $(1-2m/r_0)^{-1}$ if the 
observer is placed at $r_0$ above the radius $r_1$; so the light 
will appear to him red-shifted if received from $r<r_0$ and 
blue-shifted if received from $r>r_0$.\pni 
Coming back to the line element (4), we want to replace the 
radial coordinate $r$ with an angular coordinate $\psi$; because 
of the covariance of Einstein's equations there are infinite ways
to accomplish the replacement. We choose to define an angular 
coordinate $\psi$ given by 
\begin{equation}
r = \dfrac{2m}{\cos^2\psi} 
\hspace{4cm} -\, \dfrac{\pi}{2} \leq \psi \leq \dfrac{\pi}{2} 
\end{equation}
when $r > 2m$, and analytically continued to
\begin{equation}
r = \dfrac{2m}{\cosh^2\psi}
\hspace{3.7cm} -\, \infty < \psi <  \infty 
\end{equation}
when $r < 2m$.
The line element (4) becomes, in the region $r > 2m$ 
\begin{equation}
ds^2 = \dfrac{16 m^2}{\cos^6 \psi}\, d\psi^2 + \dfrac{4 
m^2}{\cos^4\psi}\, d\Omega^2 -\dfrac{\sin^2\psi}{\sin^2\psi_0}\, 
dt^2
\end{equation}
Here the event horizon is placed at $\psi = 0$, while infinity is
reached at $\psi = \pm \, \pi/2$.  The metrical relations in the 
surface $t$ = constant, $\vartheta = \pi/2$ are illustrated by 
means of the surface of revolution $f(r) = \sqrt{8m(r-2m)}$ 
(remember the representation of the Flamm's paraboloid with the 
Einstein-Rosen bridge).
In the extended region $r < 2m$ one has instead the line 
element 
\begin{equation}
ds^2 = -\, \dfrac{16 m^2}{\cosh^6 \psi}\, d\psi^2 + \dfrac{4 
m^2}{\cosh^4\psi}\, d\Omega^2 
+ \dfrac{\sinh^2\psi}{\sinh^2\psi_0}\, dt^2
\end{equation}
which describes the interior of a black hole joined to the 
exterior by the event horizon placed at $\psi = 0$.
It is worth noticing that the introduction of the $\psi$ 
coordinate provides a division of the maximally 
extended Schwarzschild spacetime in four regions with two 
singularities corresponding to an equal gravitational mass, just 
as described by Kruskal-Szekeres coordinates. These 
singularities are placed at $\psi = \pm \infty$, being now 
$\psi$ a time coordinate. 
\pni\emph{The interior Schwarzschild solution}:\quad 
The gravitational field inside a celestial body, 
say a star, modelled on an ideal fluid medium with 
energy-momentum tensor \begin{equation} T_{\mu\nu} = (\rho + p) 
u_\mu u_\nu + p g_{\mu \nu} \end{equation} is given, for static 
distribution of matter and pressure and moreover under the 
hypotheses of spherical symmetry and constant mass density, by 
\begin{equation}
ds^2 = \dfrac{dr^2}{1-\dfrac{r^2}{R^2}} + r^2 d\Omega^2 - 
\left[ A - B\,\sqrt{1-\dfrac{r^2}{R^2}}\right]^2\, dt^2
\end{equation}
Here  $A$ and $B$ 
are integration constants to be determined by the matching 
conditions. We use this simple and rather unrealistic solution as
a toy model uniquely to illustrate the employ of the new angular 
coordinate. If we now define the angular coordinate $\chi$ as
\begin{equation}
\dfrac{r}{R} \equiv \sin\chi \hspace{6cm} 0 \leq \chi \leq \pi 
\end{equation}
the line element (10) becomes
\begin{equation}
ds^2 = R^2\, (d\chi^2 + \sin^2\chi d\Omega^2) - [A - 
B \cos\chi ]^2\, dt^2
\end{equation}
From Einstein's equations the pressure $p$ and the mass density 
$\rho$ are 
\begin{equation}
p = \dfrac{1}{8\pi R^2}\, \left[ \dfrac{3B 
\cos\chi-A}{A-B \cos\chi}\right]\ , \hspace{2.6cm} \rho = 
\dfrac{3}{8 \pi R^2} 
\end{equation}
In formulating the matching conditions to connect the exterior 
and interior Schwarzschild solutions, continuity of the metric 
and its derivatives are to be taken into account. However in
our simple example we rest on physical plausibility 
considerations, so we require that the metric is continuous 
for $\sin \chi_1=~\dfrac{r_1}{R}$, where $r_1$ is the radius 
of the body, that the pressure $p$ vanishes on its surface and 
that the observer is in the interior at an angle $\chi_0$. \pni 
As a result one obtains 
\begin{equation} \hspace{-1cm}
\sin \chi_1 = \left(\dfrac{2 m}{R}\right)^{1/3}\ 
,\hspace{3mm} A = \dfrac{3\cos\chi_1}{3\cos\chi_1 - 
\cos\chi_0} \ ,\hspace{3mm}  B = \dfrac{1}{3\cos\chi_1 - 
\cos\chi_0}
\end{equation}
where $m$ is the gravitational mass.
The line element (10) can now be written
\begin{equation}
ds^2 =  R^2 \, (d\chi^2 + \sin^2\chi 
d\Omega^2) - \left[\dfrac{3 \cos\chi_1 - \cos\chi}{3 \cos\chi_1 
- \cos\chi_0} \right]^2\, dt^2
\end{equation}
So the observer receives the frequency of light red-shifted when 
coming from inside and blue-shifted when coming from outside.
The matching to the exterior solution requires that
\begin{equation}
N^2 = \left[\dfrac{2\cos\chi_1}{3\cos\chi_1-\cos\chi_0}
\right]^2\, \left(1 - \dfrac{2m}{R\sin\chi_1}\right)^{-1} 
\end{equation}
If the observer is at the exterior the previous values of $A$ and
$B$ change accordingly.
The pressure becomes 
\begin{equation}
p = \dfrac{3}{8\pi R^2}\, \left[ \dfrac{\cos\chi - \cos\chi_1}{3 
\cos\chi_1 - \cos\chi}\right]
\end{equation}
and is obviously observer independent.
Because of definition (11) two cases are now to be considered,
depending whether for a given value of $r_1$ one chooses $\chi_1 
< \pi/2$ or $\chi_1 > \pi/2$. In the former case, while the mass 
density $\rho$ is constant, the pressure $p$, which is zero at 
the surface, increases inwards; the solution is non singular as 
long as $p$ is finite. At $r = 0$ where $p$ takes its maximum 
value, this is only possible for $\chi_1^{(1)} < \arccos\, (1/3) 
\approx 0.39 \pi$, that is, as known [2], for $r_1/(2m) > 9/8$. 
In the latter case, the pressure $p$ takes negative values in the
interior, and the solution is non singular at $r=0$ for 
$\chi_1^{(2)} > \pi/2$. In both 
cases, the weak energy condition
\begin{equation}
\rho \geq 0 \ , \hspace{1cm} \rho + p \geq 0
\end{equation}
is always satisfied. We would also point out that while the 
surface area $ S = 4 \pi R^2 \sin^2\chi_1$ is the same in the
two cases, independently of the choice made for $\chi_1$, things 
are different in calculating volumes, given by the formula
\begin{equation}
V = 4 \pi R^3\, \int_{0}^{\chi_{1}} \sin^2 
\chi \, d\chi = \pi R^3 (2 \chi_1 - \sin 
2\chi_1)
\end{equation}
To make an example let us consider two bodies having the 
same gravitational mass but different values of 
$\chi_1$ given respectively by $\chi_1^{(1)}$ and $\chi_1^{(2)} =
\pi - \chi_1^{(1)}$ (and so the same value of 
$\sin\chi_1$). The ratio $V^{(2)}/V^{(1)}$ of their volumes is
\begin{equation}
\dfrac{ V^{(2)}}{V^{(1)}} = \dfrac{2(\pi - \chi_1^{(1)}) + \sin 
2\chi_1^{(1)}}{\chi_1^{(1)} - \sin2\chi_1^{(1)}} 
\end{equation}
Therefore while the volume $V^{(1)}$ encloses a star whose matter 
is endowed by the usual properties ($\rho > 0,\ p>0$), the volume 
$V^{(2)}$ may be so large to be considered as a 
``quasi-universe'', so named because it is an universe deprived 
of a spherical void, containing  matter with unusual 
properties ($\rho > 0,\ p < 0$); we do not call such a matter 
exotic, because it satisfies the weak energy condition and so 
also the null energy condition [3]. The connection between a body
and a quasi-universe through a suitable part of the Flamm 
paraboloid is schematically represented in Figure 1. A different 
possibility is shown in Figure 2 where now two 
quasi-universes are joined through an Einstein-Rosen bridge 
(with throat at $\psi = 0$) which can be renamed ``extreme 
wormhole''; here the matching  conditions to be fulfilled for the
second junction are the same already seen for the first, 
analogous quantities being now renamed with the same letter 
primed. Because the throat is in the vacuum, the null energy 
condition is not violated; so, according to the Morris-Thorne 
analysis [5] it is not seen as traversable by an observer placed
in a fixed forwarding station. The Einstein-Rosen bridge (or 
extreme wormhole) can also be considered as a limiting case, when
the post-Newtonian parameter $\gamma \to 1^+$, of the 
corresponding Brans-Dicke solution [6]. Finally, because of the 
necessary equality of the gravitational masses in the three 
joined solutions, one obtains the following relation between the 
densities of the two quasi-universes $\rho$ and~$\rho_{1}^{'}$: 
\begin{equation} \dfrac{\rho}{\rho_{1}^{'}} = 
\left(\dfrac{\sin\chi_1}{\sin\chi_{1}^{'}}\right)^2 
\end{equation} \pni\emph{Conclusions}: \quad 
The metrics corresponding to the exterior and interior 
\linebreak Schwarzschild solutions have 
been rewritten replacing the usual radial coordinate with an 
angular one. With respect to the exterior solution, it covers 
four different regions of the space-time. With respect to the
interior solution, it has been extended from the case $\chi < 
\pi/2$ (first-type solution) to the case $\chi > \pi/2$ of a 
quasi-universe (second-type solution). A second-type solution
can be joined either to a first-type or to a second-type solution
respectively through a suitable part of the Flamm's paraboloid 
or through a particular Einstein-Rosen bridge (extreme 
wormhole) provided the gravitational masses are equal. 
\pni Let us now consider Equations (7) and (8) in the limiting 
case when the exterior Schwarzschild solution goes over all the 
remaining space (asymptotic flatness). It is our opinion that 
the following unions ($\bigcup$) of two of the four regions 
- named hereafter $\mathit{I,II,III,IV}$ according to the 
customary nomenclature [4] - of the Kruskal-Szekeres diagram  
give rise to distinct solutions: 
\pni 1) $\mathit{I}\bigcup\mathit{II}$ : there is 
a singularity corresponding to a gravitational mass $m$ at $\psi 
= -\, \infty$  and a quasi-universe of 
gravitational mass $m$ and density $\rho = 0$ at the 
boundary $\psi = \pi/2$. The two regions are separated at $\psi =
0$ by an event horizon.  \pni 2) $\mathit{III}\bigcup\mathit{IV}$ 
: there is a singularity corresponding to a gravitational mass 
$m$ at $\psi = + \infty$  and a quasi-universe
of gravitational mass $m$ and density $\rho = 0$ at the 
boundary $\psi = -\,\pi/2$. The two regions are separated at 
$\psi = 0$ by an event horizon.  \pni 3) 
$\mathit{I}\bigcup\mathit{III}$ : there are two 
quasi-universes with gravitational mass $m$ and density $\rho
= 0$ at the boundaries $\psi = \pi/2$ and $\psi = -\, \pi/2$, 
connected by an extreme wormhole. 
\pni 4) $\mathit{II}\bigcup\mathit{IV}$ : the universe consists 
of two equal masses placed respectively at $\psi = -\, \infty$ 
and at $\psi = + \infty$ with a cosmological horizon at $\psi = 
0$. \pni The Penrose diagram for the maximally extended 
Schwarzschild spacetime is a representation of the set of the 
four solutions. \pni More in general, one could consider 
expanding quasi-universes, which are universes with cavities 
[7],[8],[9],[10]. Inside one of such voids there is a body whose 
inertial mass is, by the equivalence principle, equal to its own 
gravitational mass and consequently, broadening the above 
considerations,  also to the gravitational mass of the 
quasi-universe. Otherwise stated, the inertial mass of a body 
could be equal to the gravitational mass of its surrounding 
quasi-universe. While for a closed universe the angular momentum 
is ``undefined and undefinable'' [11], this is not true for a 
quasi-universe, and therefore it would be interesting to 
investigate whether its angular momentum is equal and opposite to
that of the body in the cavity. As an example, the Kerr-dS 
solution does represent an universe containing two equal masses 
with two equal and opposite angular momenta [12]. If this turns 
out to be the general case and if one remembers what stated 
before about the equality of the inertial mass of the body to the
gravitational mass of the quasi-universe, it will prove 
attractive to search a link between these facts and  Mach's 
principle. 
\newpage 
\begin{flushleft}
\Large\textbf {Figure captions}
\end{flushleft}
Figure 1: The connection between a body and a quasi-universe.
\pni Figure 2: The connection between two quasi-universes.
\newpage
\begin{figure}
    \centering
   \includegraphics[scale=0.8]{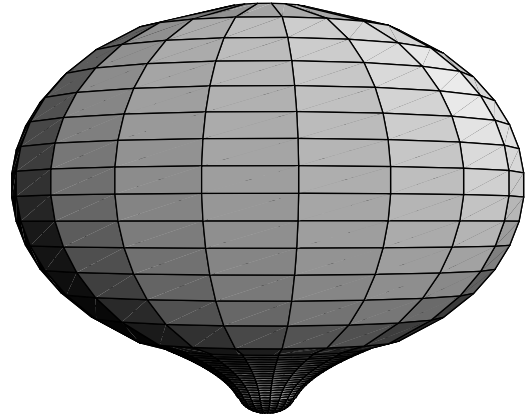}
    \caption{The connection between a body and a quasi-universe.}
\end{figure} \nopagebreak
\begin{figure}
    \centering
   \includegraphics[scale=0.8]{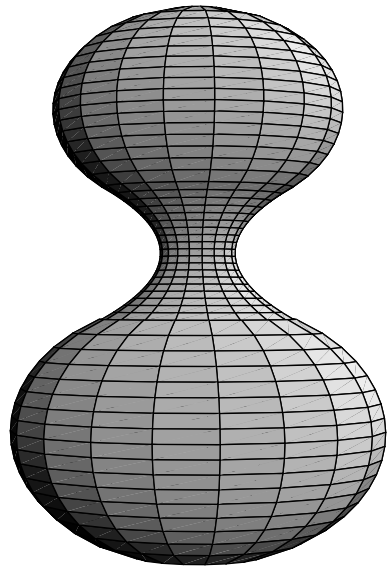}
    \caption{The connection between two quasi-universes.} 
\end{figure}
\end{document}